\documentclass[10pt,letterpaper]{article}
\usepackage{opex3}
\usepackage{epstopdf}
\usepackage{amsmath}
\usepackage{url}
\usepackage{cite}
\usepackage{color}

\begin{document}

\setlength{\tabcolsep}{15pt}

%%%%%%%%%%%%%%%%%% title page information %%%%%%%%%%%%%%%%%%
\title{Photon counting compressive depth mapping}

\author{Gregory A. Howland,$^{1,*}$ Daniel J. Lum,$^{1}$ Matthew
  R. Ware$^{2}$ and John C. Howell$^{1}$}

\address{$^{1}$University of Rochester Department of Physics and
  Astronomy, 500 Wilson Blvd, Rochester, NY 14618, USA\\

  $^{2}$Department of Physics, Campus Box 4560, Illinois State
  University, Normal, IL 61790, USA}

\email{*ghowland@pas.rochester.edu} 

\begin{abstract}
  We demonstrate a compressed sensing, photon counting lidar system
  based on the single-pixel camera. Our technique recovers both depth
  and intensity maps from a single under-sampled set of incoherent,
  linear projections of a scene of interest at ultra-low light levels
  around $0.5$ picowatts. Only two-dimensional reconstructions are
  required to image a three-dimensional scene. We demonstrate
  intensity imaging and depth mapping at $256\times 256$ pixel
  transverse resolution with acquisition times as short as $3$
  seconds. We also show novelty filtering, reconstructing only the
  difference between two instances of a scene. Finally, we acquire
  $32\times 32$ pixel real-time video for three-dimensional object
  tracking at $14$ frames-per-second.
\end{abstract}

%(110.6880) Three-dimensional image acquisition
%(110.1758) Computational imaging 
%(280.3640) Lidar


% References
%\bibliography{refs}
%\bibliographystyle{osajnl}

\section{Introduction}

High-resolution three-dimensional imaging, particularly using
time-of-flight (TOF) lidar, is very desirable at ultra-low light
levels. Such weakly-illuminated systems are safer, require less power,
and are more difficult to intercept than their high-power
equivalents. Potential applications include large- and small-scale
surface mapping, target recognition, object tracking, machine vision,
eye-safe ranging, and atmospheric sensing
\cite{rioux:2001,mallet:2009,hussman:2009,foix:2011,schwarz:2010}.

In a TOF lidar system, a scene is illuminated by laser pulses. The
pulses reflect off targets in the scene and return to a
detector. Correlating the returning signal with the outgoing pulse
train provides the TOF, which is converted to distance-of-flight
(DOF). Transverse spatial resolution can be obtained either by
scanning through the scene pixel-by-pixel or by using a detector
array, such as a time-resolving CCD.

To achieve low-light capability, many recent efforts use
photon-counting detectors for the detection element(s). Such detectors
provide shot-noise limited detection with single-photon sensitivity
and sub-ns timing. Examples include established technologies such as
photo-multiplier tubes and geiger-mode avalanche photo-diodes (APDs),
as well as more experimental devices such as superconducting nano-wires
\cite{mccarthy:2013}.

While they provide exceptional sensitivity and noise performance,
photon-counting systems are challenging to scale to high transverse
resolution. Systems that scan a single-element detector benefit from
mature technology, but suffer from acquisition times linearly
proportional to the spatial resolution. Significant per-pixel dwell
times limit real-time performance to low resolution.

Alternately, researchers have constructed photon counting detector
arrays, such as that used by MIT Lincoln Labs JIGSAW
\cite{richard:2005, heinrichs:2007} and its successors
\cite{entwistle:2012}.  Fabrication difficulties limit sensor sizes to
a current-best resolution of $32\times 128$ pixels, with $32\times 32$
pixels available commercially \cite{itzler:2011}. These sensors suffer
from high dark count rates, pixel cross-talk, and significant readout
noise. For TOF ranging, each pixel must be simultaneously correlated
with the pulse train, a resource intensive process.

The goal is to overcome these challenges without increasing system
cost and complexity. A promising option uses a standard,
single-element, photon counting detector as the detection element in a
compressive sensing, single-pixel camera \cite{duarte:2008}. The
single-pixel camera leverages a scene's compressibility to recover an
image from an under-sampled series of incoherent projections. This is
accomplished by imaging the scene onto a sequence of psuedo-random
patterns placed on a spatial light modulator (SLM) or digital
micro-mirror device (DMD). The inner-product of the scene and pattern
is measured by a single-element, ``bucket'' detector. Optimization
routines then recover the image from many fewer projections than
pixels, often less than 10 percent.

The single-pixel camera can be adapted for TOF depth mapping by
switching to active, pulsed illumination.  Standard CS techniques
require measurements to be linear projections of the signal of
interest. Unfortunately, in a TOF single-pixel camera, the depth map
is non-linearly related to the acquired measurements, as they contain
both depth and intensity information
\cite{sarkis:2009,Kirmani:2011}. Current approaches mitigate or
altogether avoid this issue in various ways.  In 2011, we presented a
proof-of-principle lidar camera where CS is used only to acquire
transverse information \cite{howland:2011}. Range is determined by
gating on a TOF of interest. This reduces to the conventional
single-pixel camera problem, but requires separate reconstructions for
each depth of interest. Li et. al. present a similar, gated system
(GVLICS) that recovers 3D images \cite{li:2012}. Kirmani
et. al. \cite{Kirmani:2011} introduce a more sophisticated protocol
(CoDAC) that incorporates parametric signal processing and a
reconstruction process exploiting sparsity directly in the depth
map. This system uses intermediate optimization steps to keep the
problem linear.

CS can also be applied directly in the time-domain, and has been
proposed for non-spatially resolving lidar systems
\cite{babbitt:2011}. In principle, one could combine this with
transverse CS for a full 3D voxel reconstruction. This signal is
extremely high dimensional and presents substantial measurement and
reconstruction challenges.

In this manuscript, we present an alternative single-pixel camera
adaptation for photon-counting depth mapping and intensity
imaging. Rather than recover the depth map directly, we separately
recover an intensity map and an intensity $\times$ depth map, where
the depth map is simply their ratio. Linear projections of these
signals are easily acquired using a TOF single-pixel camera with a
photon counting detection element, with available light levels around
$0.5$ picowatts. Our approach is straightforward and uses standard CS
formulations and solvers. We demonstrate imaging at resolutions up to
$256\times 256$ pixels with practical acquisition times as low as $3$
seconds. We also show novelty filtering; finding the difference
between two instances of a scene without recovering the full
scenes. Finally, we demonstrate video acquisition and object tracking
at $32\times 32$ pixel resolution and $14$ frames per second.

\section{Compressive Sensing}

\subsection{Introduction to CS}

Compressive sensing \cite{donoho:2006} is a measurement technique that
uses optimization to recover a sparsely represented, $n$-dimensional
signal $X$ from $m<n$ measurements. CS exploits a signal's
compressibility to require fewer measurements than the Nyquist
limit. The detection process can be modeled by interacting $X$ with a
$m\times n$ sensing matrix $A$ such that
\begin{equation}
  Y = AX + \Gamma,
  \label{eq:meas}
\end{equation}
where $Y$ is a $m$-dimensional vector of measurements and $\Gamma$ is
a $m$-dimensional noise vector. Because $m<n$, and rows of $A$ are not
necessarily linearly independent, a given set of measurements does not
specify a unique signal.

To recover the correct signal, CS formulates an objective function to
select the sparest $X$ consistent with the measurements:
\begin{equation}
  \min_X \frac{1}{2}||Y-AX||^{2}_{2}+\tau g(X),  
  \label{eq:objective}
\end{equation}
where $\tau$ is a scaling constant between penalties. The first
penalty is a least-squares term that ensures the signal matches the
measured values. The second penalty, $g(X)$, is a sparsity-promoting
regularization function. Typical $g(X)$ include the $\ell_1$ norm of
$X$ or the total variation of $X$. For a $k$-sparse signal (only $k$
significant elements), exact reconstruction is possible for $m \propto
k\log(n/k)$ measurements. In practice, $m$ is often only a few percent
of $n$. For minimum $m$, the sensing matrix must be incoherent with
the sparse basis, with the counter-intuitive result that random,
binary sensing vectors work well \cite{candes:2007}.

\subsection{Single-pixel Camera}

The quintessential application of CS is the single-pixel camera
\cite{duarte:2008}. Duarte et. al. cleverly implemented an incoherent
sensing matrix by imaging a scene onto a digital micro-mirror device
(DMD) of the sort found in digital movie projectors. The DMD is an
array individually addressable mirrors that can be selectively
oriented to reflect light either to a single-pixel ``bucket'' detector
or out of the system. The measured intensity yields the inner-product
of a pattern placed on the DMD with the scene of interest. To perform
the set of measurements described in Eq. (\ref{eq:meas}), pseudorandom
binary patterns are placed sequentially on the DMD, where each pattern
represents a row of sensing matrix $A$. Elements of $Y$ are the
measured intensities for corresponding patterns. An image of the scene
is then recovered by an algorithm that solves Eq. (\ref{eq:objective}).

CS and the single-pixel camera sparked a mini-revolution in sensing as
researchers found their ideas could be easily and directly applied to
common problems across a breadth of fields from magnetic resonance
imaging \cite{lustig:2007} to radio astronomy \cite{bobin:2008} to
quantum entanglement characterization \cite{flammia:2012,
  howland:2013}.

\section{Compressive Depth Mapping}

\subsection{Adapting SPC for Depth Mapping}

\begin{figure}[h]
  \begin{centering}
    \includegraphics[scale=0.20]{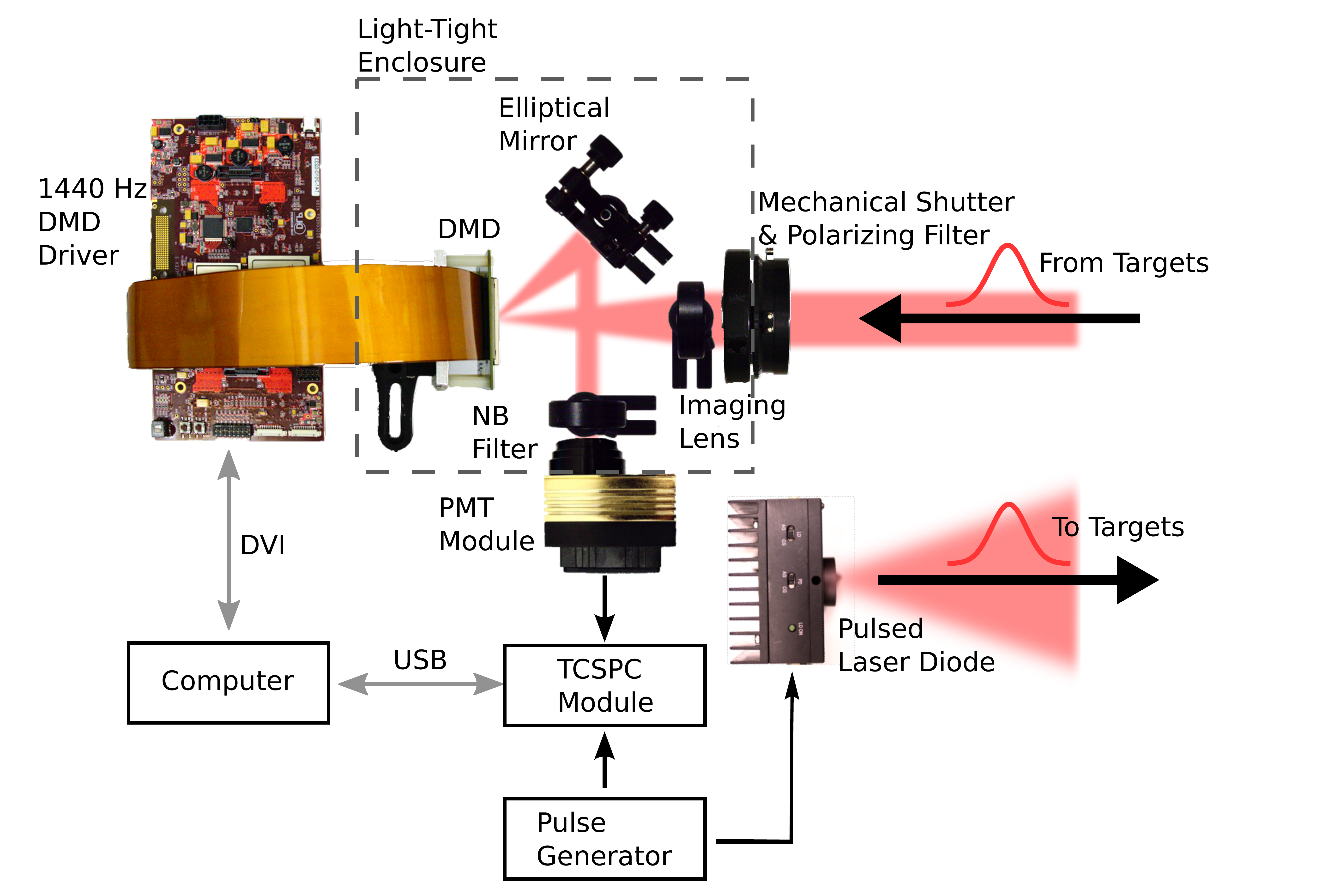}
    \caption{\textbf{Experimental Setup} A 780 nm, 10 Mhz, 2 ns pulsed
      laser diode flood illuminates a scene containing targets at
      different depths. Returning pulses are imaged onto a DMD array
      with resolution up to $256\times 256$ pixels. A polarizer
      prevents flares from specular reflection. Light reflecting off
      DMD ``on'' pixels is directed through a narrow-band filter to a
      single-photon sensitive PMT module that produces TTL pulses.
      Typical count rates are about $2$ million photons per second. A
      TCSPC module time-correlates photon arrivals with the outgoing
      pulse train to determine a TOF for each detected photon. A
      series of psuedorandom, binary patterns consisting of randomly
      permuted, zero-shifted Hadamard patterns are placed on the DMD
      with per-pattern dwell times as short as $1/1440$ sec. These
      implement an incoherent sensing matrix. For each pattern, the
      number of photon arrivals and their total TOF is recorded. Our
      protocol is then used to reconstruct the intensity image and the
      depth map.}
    \label{fig:setup}
  \end{centering}
\end{figure}

The single-pixel camera can be adapted for depth mapping at very low
light levels by switching to active, pulsed illumination and using a
photon-counting detection element. Our setup is given in
Fig. \ref{fig:setup}. A scene containing targets at different depths
is flood illuminated by a pulsed laser diode. The scene is imaged onto
a DMD which implements an incoherent sensing matrix. Light from DMD
``on'' pixels is redirected to a photo-multiplier module that detects
individual photon arrivals. A time-correlated single-photon counting
module (TCSPC) correlates photon arrivals with the outgoing pulses to
find each photon's TOF. Full experimental specifications are given in
Table \ref{tab:specifications}.

Recovery of a two-dimensional, intensity only image $X_I$ (a
gray-scale image) is identical to the normal single-pixel
camera. Patterns from the sensing matrix are placed sequentially on
the DMD. For each pattern, the number of detected photons is recorded
to obtain a measurement vector $Y_I$. Eq. (\ref{eq:objective}) is then
used to find the intensity image $X_I$.

Unfortunately, recovering a depth map $X_D$ from the TOF information
is not straightforward because the measurements are nonlinear in
$X_D$. Consider photon arrivals during one pattern. Each photon has a
specific TOF, but is only spatially localized to within the sensing
pattern. It is possible, and likely, that multiple photons will arrive
from the same location in the scene. Individual detection events
therefore contain information about both intensity and depth.

Consider the signal $X_Q$ made up of the element-wise product $X_Q =
X_I.X_D$. Unlike $X_D$, $X_Q$ can be linearly sampled by summing the
TOF of each photon detected during the measurement. This can be seen
by expanding this total TOF sum over the contribution from each pixel
in the scene, such that
\begin{equation}
  Y_{Q} = A_{}X_Q = \left\{C\sum\limits_{j=1}^NA_{ij} \eta_j T_j \right\} \text{ for } i=1 \text{ to } m,
  \label{eq:qexp}
\end{equation}
where $i$ is an index over sensing patterns and $j$ is an index over
all DMD pixels. $A_{ij}$ is the DMD status ($0$ or $1$) of pixel $j$
during the $i^{\text{th}}$ pattern, $\eta_j$ is the number of photons
reaching pixel $j$ during the measurement, and $T_j$ is the TOF of a
photon arriving at pixel $j$. $C$ is a constant factor converting
photon number to intensity and TOF to depth.  Each pixel where $A_{ij}
= 1$ (the DMD mirror is ``on'') contributes $\eta_{j}T_{j}$ to the TOF
sum for pattern $i$. This is a product of an intensity (photon number)
and a depth (TOF). $X_Q$ is therefore equal to $C\eta .T$. Because
Eq. (\ref{eq:qexp}) takes the form of Eq. (\ref{eq:meas}),
Eq. (\ref{eq:objective}) is suitable for recovering $X_Q$.

The depth map is simply the element-wise division of $X_Q$ by
$X_I$. Note that $Y_I$ and $Y_Q$ are acquired simultaneously from the
same signal; $Y_I$ represents that number of photon arrivals for each
pattern and $Y_Q$ is the sum of photon TOF for each pattern. The only
increase in complexity is that two optimization problems must now be
solved, but these are the well understood linear problems
characteristic of CS.

\subsection{Protocol}

A slightly more sophisticated approach including noise reduction
vastly improves practical performance. The protocol we use for depth
map recovery is as follows:

\begin{enumerate}
\item Acquire measurement vectors $Y_Q = AX_Q$ and $Y_I = AX_I$, where
  $X_Q=X_I.X_D$.
\item Use sparsity maximization routine (Eq. (\ref{eq:objective})) on
  $Y_Q$ to recover noisy $\bar{X_Q}$.
\item Apply hard thresholding to $\bar{X_Q}$. The subset of non-zero
  coefficients in a sparse representation of $\bar{X_Q}$ now form an
  over-determined, dense recovery problem.
\item Perform a least squares debiasing routine on non-zero
  coefficients of $\bar{X_Q}$ in the sparse representation to find
  their correct values and recover $X_Q$.
\item Take significant coefficients of $X_I$ to be identical to
  $X_Q$. Perform the same least squares debiasing on these
  coefficients of $X_I$.
\item Recover $X_D$ by taking $\text{Nz}(X_I).X_Q./X_I$, where $Nz(x)
  = 1$ for non-zero $x$ and $0$ otherwise.
\item (optional) In the case of very noisy measurements, perform
  masking on $X_D$ and $X_I$.
\end{enumerate}

First, we acquire measurement vectors $Y_I$ and $Y_Q$ as previously
described. Rather than independently solve Eq. (\ref{eq:objective}) for
both measurement vectors, we only perform sparsity maximization on
$Y_Q$. In practice, solvers for Eq. (\ref{eq:objective}) tend to be more
effective at determining significant coefficients than finding their
correct values, particularly given noisy measurements
\cite{figueiredo:2007}. We therefore rely on sparsity maximization
only to determine which coefficients are significant. We subject the
noisy, recovered $\bar{X_Q}$ to uniform, hard thresholding
\cite{donoho:1994:2} in a sparse representation, which forces the
majority of the coefficients to zero. The sparse basis is typically
wavelets, but may be the pixel basis for simple images.

The subset of coefficients remaining after thresholding is considered
significant. If only these coefficients are considered, the problem is
now over-determined and a least-squares debiasing routine is applied
to find their correct values and yield the signal $X_Q$
\cite{figueiredo:2007}. We assume the significant coefficients for
$X_I$ are the same as for $X_Q$, and apply least-squares debiasing to
$X_I$ as well. Finally, we recover the depth map $X_D$ by taking
\begin{equation}
  X_D = \text{Nz}(X_I).X_Q./|X_I|
\label{eq:xd}
\end{equation}
where $\text{Nz}(p) = 1$ for nonzero $p$ and $0$ otherwise. This
prevents a divide-by-zero situation when an element of $X_I = 0$,
meaning no light came from that location.

For very noisy measurements, the sparsity maximization in steps 2-3
gives accurate outlines for targets in the scene, but poorly recovers
pixel values. After the least squares debiasing in steps 4 and 5,
values are more accurate, but the outlines can become distorted. A
best-of-both-worlds solution is to threshold the result of step 4 in
the pixel basis to create a binary object mask. This mask can
optionally be applied to the final $X_I$ and $X_D$ for spatial clean
up.

\section{Experimental Setup}

\begin{table}
\caption{\textbf{System Specifications}}

\renewcommand{\arraystretch}{1.5}
{\small
\begin{tabular}{l p{6cm}}
  \hline
  {\bf Parameter} & {\bf Comments}\\
  \hline
  {\bf Laser} & Laser Diode mounted in Thorlabs TCLDM9\\
  {\bf Repetition Rate} & 10 MHz\\
  {\bf Pulse Width} & $2$ ns $=$ $60$ cm\\
  {\bf Wavelength} & $780$ nm\\
  {\bf Average Output Power} & 4 mW \\
  {\bf Average Detected Power} & $0.5$ pW (2 megacounts per second)\\
  {\bf Spectral Filters} & 780/10 nm Bandpass, Optical Depth 6+\\
  {\bf DMD Device} & Digital Light Innovations D4100 with $1024\times768$ DMD, operated in binary expansion mode at 1440 frames per second  \\
  {\bf Spatial Resolution} & images: $256\times 256$ pixels \\
  & video: $32\times 32$ pixels\\
  {\bf Detector} &  Horiba TBX-900C Photomultiplier Module with $9$ mm$^2$ active area\\
  {\bf Jitter} & $<200$ ps\\
  {\bf Dark Count Rate} & $<200$ counts per second \\
  {\bf Data Acquisition} & PicoQuant PicoHarp 300 operating in time-tagging mode\\
  {\bf Imaging Lens Focal Length} & $75$ mm \\
  \hline
\end{tabular}
}
\label{tab:specifications}
\end{table}

Our experimental setup is given in Fig. \ref{fig:setup}, with
specifications given in Table \ref{tab:specifications}. Targets are
flood illuminated with $2$ ns, $780$ nm laser pulses with a $10$ MHz
repetition rate. Average illumination power is $4$ mW. The scene is
imaged onto a DMD array, where on pixels are redirected through a
$780/10$ nm bandpass filter to a photon-counting photo-muliplier
module. A sequence of $m$ sensing patterns are displayed on the DMD at
$1440$ Hz. For each pattern, the number of photon arrivals and the sum
of photon TOF is recorded. If $1/1440$ seconds is an insufficient
per-pattern dwell time $t_p$, the pattern sequence is repeated $r$
times so that $t_p=r/1440$ and the total exposure time $t$ is $t =
mt_p = mr/1440$. The average detected photon rate is $2$ million
counts per second, or $0.5$ picowatts.

Sensing vectors (patterns) are randomly selected rows of a
zero-shifted, randomly-permuted Hadamard matrix. The Hadamard
transform matrix is closely related to the Fourier transform matrix,
but entries only take real values $1$ or $-1$. The zero shifted
version sets all $-1$ values to zero. When randomly permuted, they
make practical sensing vectors because repeated calculations of $AX$
in the reconstruction algorithms can be computed via fast
transform. This is computationally efficient because the fast
transform scales logarithmically and the sensing matrix does not need
to be stored in memory \cite{li:2009}.

The sparsity promoting regularization is either the signal's
$\ell_{1}$ norm or total variation depending on the scene and exposure
time. For $\ell_{1}$ minimization, we use a gradient projection solver
\cite{figueiredo:2007}. For TV minimization, We use the TVAL3 solver
\cite{li:2009}. The sparse representation for wavelet thresholding and
least-squares debiasing is Haar Wavelets. Higher order Daubechies
wavelets were tested, but did not significantly improve results for
test scenes presented in this manuscript. They may offer improvements
for more complicated scenes in future work.

Processing times for reconstructions vary based on transverse
resolution, scene complexity, and noise. Test scenes in this
manuscript were recovered either at resolutions of $n= 256\times 256$
or $n=32\times 32$. A $256\times 256$ reconstruction consistently took
less than $5$ minutes, while a $32\times 32$ reconstruction took less
than $5$ seconds. The solvers and supporting scripts are implemented
in Matlab, with no particular emphasis on optimizing their speed. The
majority of computational resources are used for performing Fast
Hadamard Transforms and reading and writing large files. If more
efficient file formats are used, and the Fast Hadamard Transform is
computed on a GPU or DSP, reconstruction speeds approach real-time.
 
\subsection{Noise Considerations}

While an exhaustive noise analysis is beyond the scope of this
manuscript and a subject of future work, some important qualitative
aspects should be explored. There are two dominant sources of error in
the measurements of $Y_{I}$ and $Y_{Q}$. The first is simply shot
noise, a fundamental uncertainly in number of measured photons
limiting the measured SNR to $\sqrt{\eta}$ for $\eta$ detected
photons. Because incoherent measurements used for compressive sensing
use half the available light on average, the information they convey
is contained in their deviation from the mean. As a general rule, a
successful reconstruction requires the standard deviation of the
measurement vector to exceed the shot noise. Compressive sensing has
been shown to work well in the presence of shot noise in comparison to
other sensing methodologies. For more information see Refs.
\cite{duarte:2008,willett:2009,donoho:2011}.

The other primary error sources are detector dark counts and ambient
light. All photon-counting detectors have the possibility of firing
without absorbing a photon. For a particular detector, these events
occur at some rate and the timing is uniformly distributed. For a
statistically significant acquisition time, the dark counts simply add
constant values to both $Y_{Q}$ and $Y_{I}$. These values can be
easily measured by blocking the detector and observing count rates in
the absence of signal. They can then be subtracted from future
measurement vectors.

Ambient light is also uncorrelated with the pulsed illumination, so
ambient photon arrivals are uniformly distributed. If the ambient
light is spatially uniform, it adds a constant value to both $Y_{Q}$
and $Y_{I}$ because half the DMD pixels are always directed to the
detector. If the ambient light is not spatially uniform, the
contribution may be different for different sensing patterns. In
either case, $m$-dimensional ambient vectors $\mathcal{A}_{Q}$ and
$\mathcal{A}_{I}$ can be obtained by disabling the pulsed laser and
performing the measurements normally used to obtain $Y_{Q}$ and
$Y_{I}$. These can then be subtracted from $Y_{Q}$ and $Y_{I}$ to
ameliorate the ambient light. This will also remove the dark counts.

If the ambient light is spatially uniform, the addition of a second
detector which measures light from DMD ``off'' pixels can also
mitigate the ambient light. In this case, it is no longer necessary to
use zero-shifted sensing patterns, they can instead consist of values
``1'', which are directed to the ``on'' pixel detector, and ``-1'',
which are directed to the ``off'' pixel detector. The sensing vector
corresponding to this matrix is just the difference between sensing
vectors obtained by each individual detector. Because each detector
receives the same amount of ambient light, it will be subtracted out.

The best way to mitigate unwanted detection events is to avoid them
altogether. A massive advantage of single-pixel camera technology is
the ability operate at wavelengths where detector arrays are difficult
to fabricate, such as the infrared. For the active, narrowband sources
used in lidar systems, high optical-density, narrow-band filters can
reject the vast majority of unwanted light. Our system received less
than $100$ dark counts per second and approximately $1000$ counts per
second from ambient light. With over $10^6$ signal counts per second,
these negligibly effected our results and the amelioration schemes
discussed were not needed. They are likely to become important in more
real-world applications beyond the laboratory.

\section{Results}
\begin{figure}[h]
  \begin{centering}
    \includegraphics[scale=0.45]{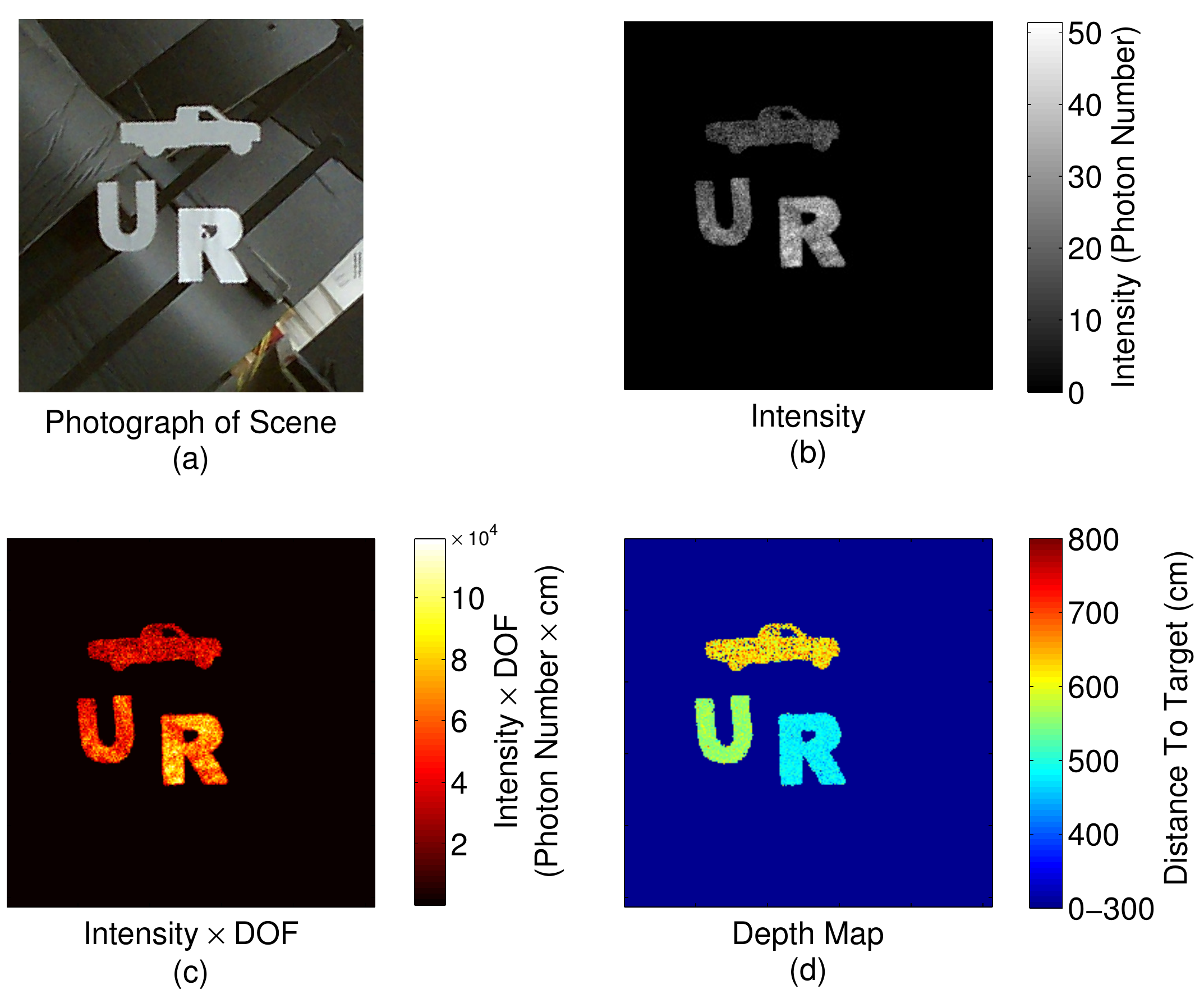}
    \caption{\textbf{Long Exposure} A scene consisting of cardboard
      cutouts (a) is imaged at $n=256\times 256$ resolution from
      $m=0.2n$ random projections with a $6.07$ minute exposure
      time. The protocol recovers a high quality intensity map (b) and
      intensity $\times$ DOF map (c). Their ratio is the depth map
      (d).}
    \label{fig:simplescene}
  \end{centering}
\end{figure}

\begin{figure}[h]
  \begin{centering}
    \includegraphics[scale=0.45]{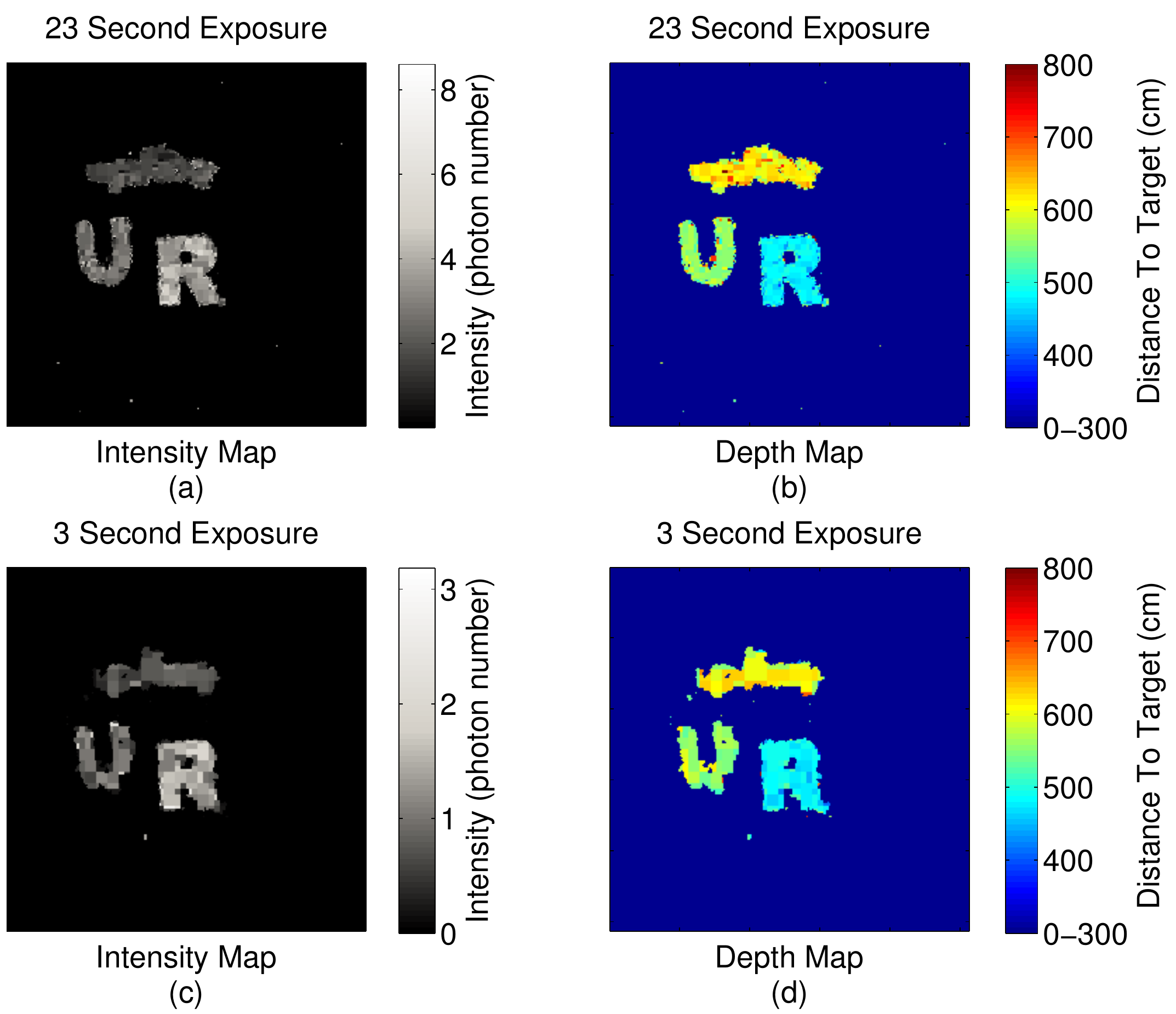}
    \caption{\textbf{Short Exposure} Scene from Fig.
      \ref{fig:simplescene} acquired with rapid acquisition times.
      The intensity and depth maps in (a) and (b) were acquired in
      $23$ seconds while (c) and (d) were acquired in only $3$
      seconds. At these exposure times, the intensity map regularly
      contains less than one photon per significant pixel, so it is
      impossible to raster scan.}
    \label{fig:fastscene}
  \end{centering}
\end{figure}

\begin{figure}[h]
  \begin{centering}
    \includegraphics[scale=0.45]{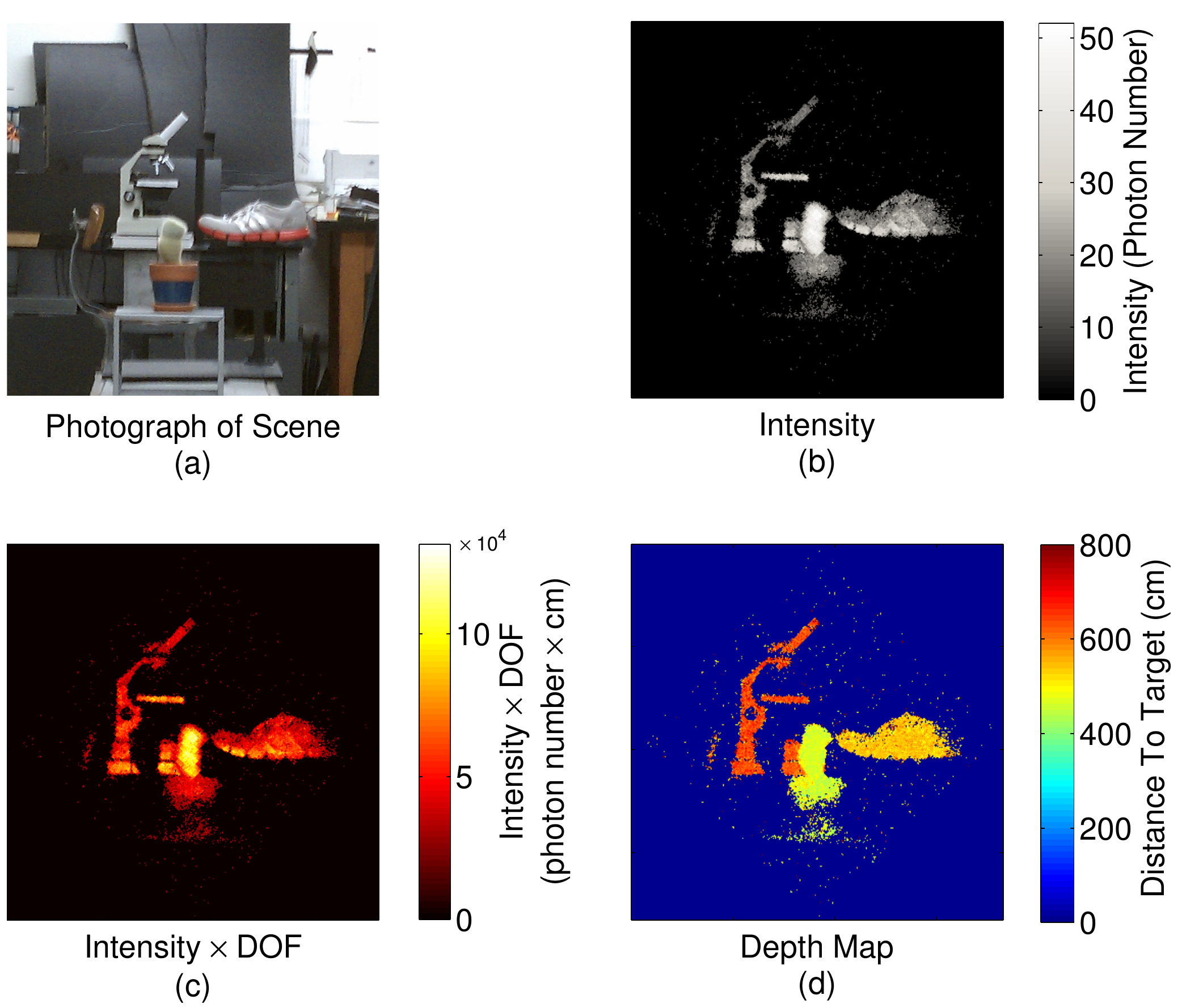}
    \caption{\textbf{Natural Scene} A scene consisting of a cactus,
      shoe, and microscope is imaged at $n=256\times 256$ pixel
      resolution with $m=0.3n$ and a per-pattern dwell time $t_p =
      50/1440$ seconds.}
    \label{fig:natural}
  \end{centering}
\end{figure}

\begin{figure}[h]
  \begin{centering}
    \includegraphics[scale=0.5]{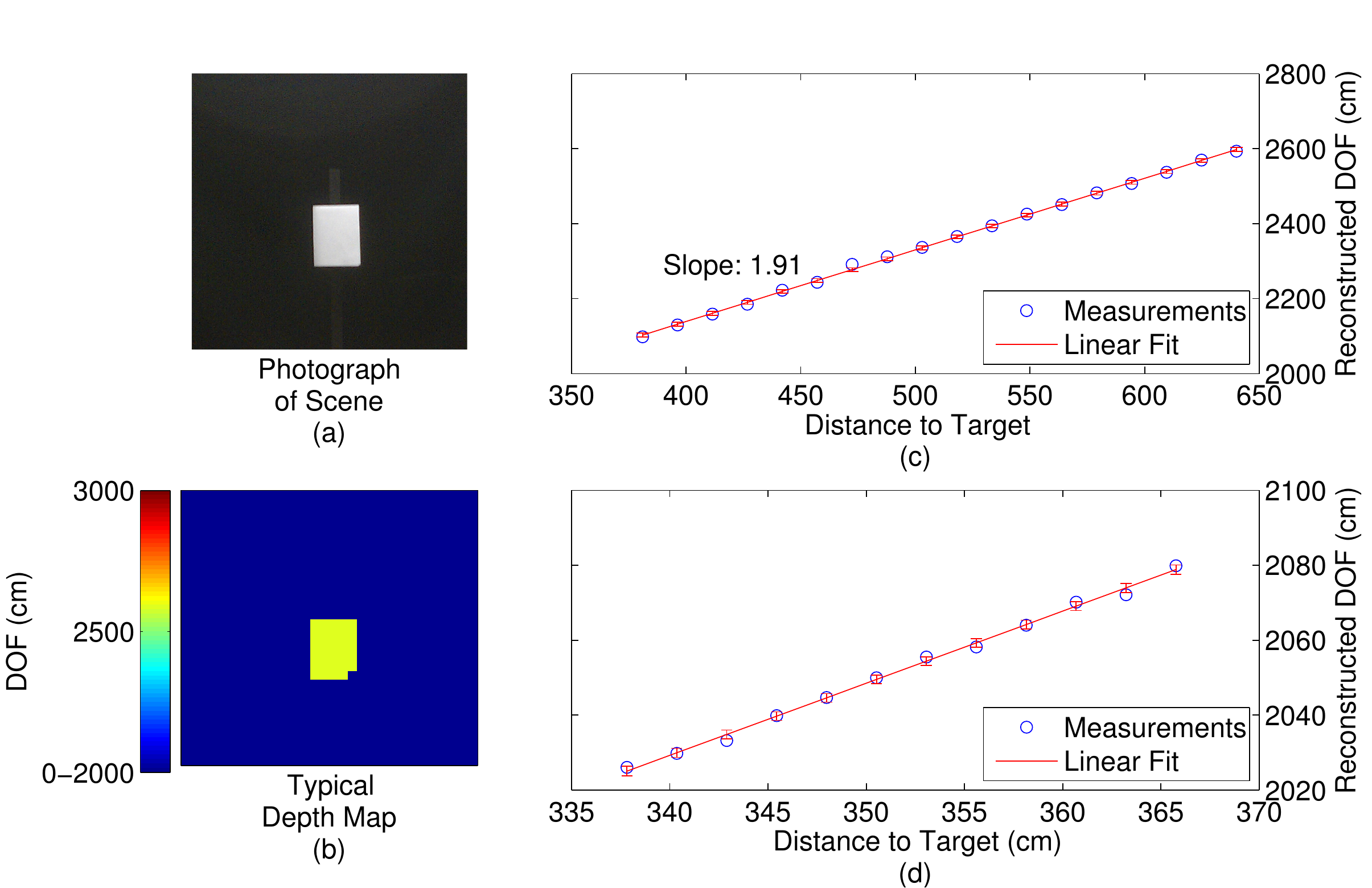}
    \caption{\textbf{Depth Calibration} Depth maps of a rectangular
      cardboard cutout (a) are acquired at $n=32\times 32$ pixel
      resolution with a typical reconstruction given in (b). The
      cutouts to-target distance was increased in increments of
      $15.52$ cm (c) and $2.54$ cm (d). Depths can accurately
      recovered to less than $2.54$ cm for this scene.}
    \label{fig:depth}
  \end{centering}
\end{figure}

\begin{figure}[h]
  \begin{centering}
    \includegraphics[scale=0.5]{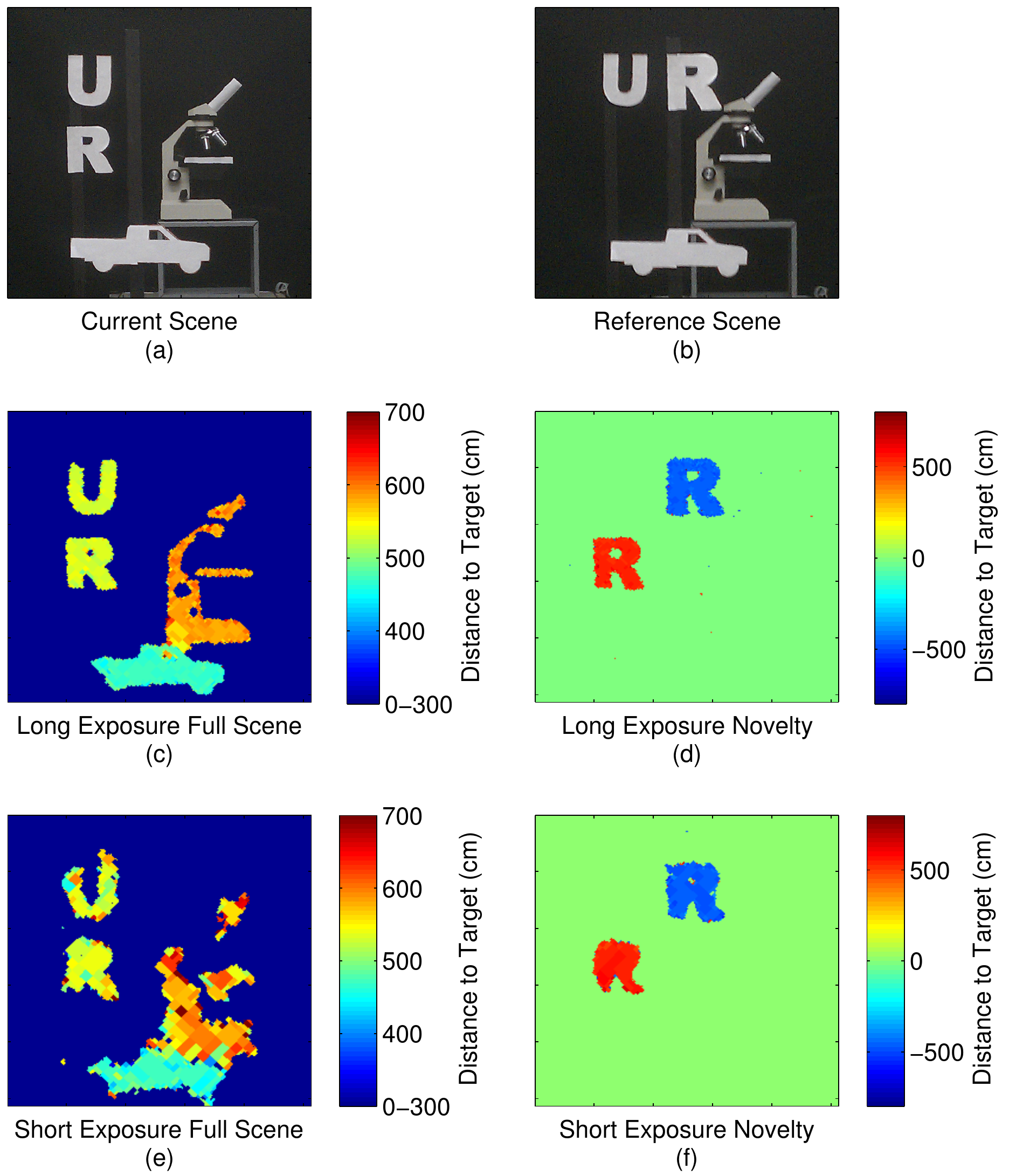}
    \caption{\textbf{Novelty Filtering} (a) and (b) give photographs
      of two instances of a scene, where the `R' has changed positions
      (including depth). (c) and (d) show high quality, long $6.07$
      minute exposure depth map reconstructions for the full current
      scene and difference image respectively. (e) and (f) show
      corresponding short $37$ second exposure depth map
      reconstructions. Negative values in the difference image
      indicate the object's former location.}
    \label{fig:novelty}
  \end{centering}
\end{figure}

\subsection{Simple Scene} % Cardboard Cutouts

The first test case is a simple scene containing cardboard cutouts of
the letters ``U'' and ``R'' as well as the outline of a truck. The
objects were placed at to-target distances of $480$ cm , $550$ cm, and
$620$ cm respectively. Fig.  \ref{fig:simplescene} shows a high
quality reconstruction with resolution $n = 256\times256$ pixels,
$m=0.2n=13108$, and $t_p=40/1440$ for an exposure time of $6.07$
minutes. The sparsity promoter was TV. Both the intensity and depth map are
accurately recovered.

Pictures of the same scene with much shorter acquisition times are
given in Fig.  \ref{fig:fastscene}. Fig. \ref{fig:fastscene}(a) and
\ref{fig:fastscene}(b) show the intensity and depth map for
$m=0.1n=6554$ and $t_p=5/1440$ seconds for a total exposure time of
$23$ seconds.  Fig.  \ref{fig:fastscene}(a) and \ref{fig:fastscene}(b)
gives results for a exposure with $m=.07n=4558$ and $t_p=1/1440$
seconds for a total exposure time of $3$ seconds.  The optional
masking process (protocol step 7) was used. While the reconstructions
are considerably noisier than the long exposure case, the objects are
recognizable and the average depths are accurate.

Note that the shortest dwell-time per pattern for the DMD is $1/1440$
sec. To raster scan at this speed for $n=256\times 256$ pixels would
require $46$ seconds, so both results in Fig.  \ref{fig:fastscene} are
recovered faster than it is possible to raster scan. If the same dwell
times were used ($t_p = 5/1440$ and $t_p = 1/1440$ seconds
respectively), the raster scan would take about $228$ seconds and $46$
seconds. Note that many significant pixels in the recovered intensity
images in Fig. \ref{fig:fastscene} have less than $1$ photon per
pixel. Therefore, even the longer raster scans would struggle to
produce a good picture because they cannot resolve fewer than $1$
photon per pixel. Our protocol is more efficient because each
incoherent projection has flux contributed from many pixels at once,
measuring half the available light on average.

A generalization of the raster scan is the basis scan, a scan through
basis vectors of an alternative signal representation (such as a
Hadamard decomposition). The hardware in most lidar systems is based
on either a detector array or raster scanning, so it is not suitable
for performing a basis scan. However, with specialized equipment, such
as the DMD in our setup, basis scanning is possible. Basis scanning
still requires $n$ measurements, so the minimum possible acquisition
time ($t_p = 1/1440$ seconds) is the same as for raster scanning, in
this case $n\times t_p = 46$ seconds. However, a basis scan using
Hadamard patterns benefits from the same increased flux per
measurement as the incoherent projections of CS, so SNR is improved
over raster scanning. Nevertheless, CS has been shown to outperform
basis scan with fewer measurements. Since the two schemes require the
same hardware, CS is preferred. For a comparison of CS with other
sensing techniques, including raster and basis scan, see
Ref. \cite{duarte:2008}.

\subsection{Natural Scene} % Microscope, etc?

A $n=256 \times 256$ pixel measurement of a more complicated scene
containing real objects is given in Fig. \ref{fig:natural}. The scene
consists of a small cactus, a shoe, and a microscope placed at
to-target distances of $465$ cm, $540$ cm, and $640$ cm
respectively. A photograph of the scene is given in
Fig. \ref{fig:natural}(a). The scene was acquired with $m=0.3n$ a
per-pattern dwell time $t_p =50/1440$ seconds for an exposure time of
$11.37$ minutes. The sparsity promoter was the $\ell_{1}$ norm in the
Haar wavelet representation.

Fig. \ref{fig:natural}(b), Fig. \ref{fig:natural}(c), and
Fig. \ref{fig:natural}(d) show the reconstructed intensity, intensity
$\times$ DOF, and depth map respectively. All objects are recognizable
and the to-target distances are accurately recovered. Noise is
slightly higher than the simpler scene of cardboard cutouts.

\subsection{Depth Calibration} % Square

The photon TOF measured by the TCSPC module is trivially converted to
DOF by multiplying by speed of light $c$. This DOF includes the round
trip distance traveled by illuminating pulses as well as delays
introduced by cables and electronics. The actual distance to target is
linearly related to the measured DOF.

To determine this relationship and probe the system's depth accuracy,
we took images of a simple square, white, cardboard cutout placed a
known distance from the camera. The cutout was sequentially moved away
from the camera in small increments and the reconstructed depths were
recorded.

Results are given in Fig. \ref{fig:depth}. Fig. \ref{fig:depth}(a)
shows a photograph of the object, with a typical reconstructed depth
map given in Fig. \ref{fig:depth}(b). Fig. \ref{fig:depth}(c) shows a
coarse grained result when the object was moved in $15.25$ cm ($6$ in)
increments over a $259$ cm ($108$ in) range. Each measurement was
performed at $32\times 32$ pixel transverse resolution with
$m=0.1n=102$. The per-pattern dwell time was $1/1440$ seconds, so each
depth map was acquired in $.07$ seconds. The pulse length was $2$ ns,
or $60$ cm.

Fig. \ref{fig:depth}(c) provides the reconstructed DOF as function of
the distance to object. The measured DOF is averaged over the
recovered object. A linear fit shows very good agreement with the
measurements, with a slope of $1.91$.  This is slightly less than the
expected $2$ (for a round-trip flight) because the result includes
cable transit times. Fit error bars represent a $95\%$ confidence
interval.

We performed an additional fine resolution set of measurements where
the object was moved in $2.54$ cm ($1$ in) increments over a $30.4$ cm
($12$ in) range.  All other parameters are identical to the coarse
measurements. Results are shown in Fig. \ref{fig:depth}(d). Despite
$60$ cm pulse lengths, depth mapping is accurate to less than $2.54$
cm. While uncertainty is slightly larger than the coarse case, the
fits agree to within the confidence interval.

\subsection{Novelty Filtering}

For many applications such as target localization, high frame-rate
video, and surveillance, it is useful to look at the difference
between realizations of a scene at different times. Such novelty
filtering removes static clutter to reveal changes in the scene. This
is traditionally accomplished by measuring the current signal
$X^{(c)}$ and subtracting from it a previously acquired reference
signal $X^{(r)}$ to find $\Delta X = X^{(c)}-X^{(r)}$.

Compressive sensing is particularly well suited for novelty filtering
because the difference signal $\Delta X$ can be directly reconstructed
\cite{cevher:2008, omar:2013}. Consider acquiring incoherent
measurement vectors for both $X^{(r)}$ and $X^{(c)}$ as in
Eq. (\ref{eq:meas}), using the same sensing matrix $A$ for both
signals. Instead of separately solving Eq. (\ref{eq:objective}) for
each signal, first difference their measurement vectors to find

\begin{align}
\Delta Y &= Y^{(c)}-Y^{(r)} \\
\Delta Y &= A \Delta X. \label{eq:diffmeas}
\end{align}

Because Eq. (\ref{eq:diffmeas}) has the same form as
Eq. (\ref{eq:meas}), the optimization routine can be performed
directly on $\Delta Y$ to obtain $\Delta X$ without ever finding
$X^{(c)}$ or $X^{(r)}$. Furthermore, the requisite number of
measurements $m$ depends only on the sparsity of $\Delta X$. For small
changes in a very cluttered scene, the change can be much sparser than
the full scene. It is often possible to recover the novelty with too
few measurements to reconstruct the full scene.

Fig. \ref{fig:novelty} gives an example of using our system for
novelty filtering in intensity and depth. A photograph of a current
scene containing several objects is shown in
\ref{fig:novelty}(a). Fig. \ref{fig:novelty}(b) photographs a prior
reference scene. In the current scene, the cardboard cutout `R' has
been moved, both transversely and from a range of $480$ cm to a range
of $560$ cm.

Fig. \ref{fig:novelty}(c) and \ref{fig:novelty}(d) give long exposure
reconstructions of the full current depth map and the difference depth
map respectively. Exposure parameters were $n=256\times 256$ pixels,
$m=0.2n$, and $t_p=40/1440$ sec for an exposure time $t=6.07$
minutes. The sparsity promoter was TV. The difference reconstruction
contains only the `R' object that moved. Note that there are two
copies of the `R'. One is negative, indicating the objects former
position, and one is positive, indicating the objects new position.

Fig. \ref{fig:novelty}(e) and \ref{fig:novelty}(f) show short exposure
reconstructions for the current scene and the difference image. The
number of measurements was reduced to $m=0.02n=1311$ with
$t_p=40/1440$ for an exposure time of $37$ seconds. Masking was
performed. For this acquisition time, the static clutter is very
poorly imaged and is difficult to recognize in the full scene. The
difference image effectively removes the clutter and gives a
recognizable reconstruction of the novelty. Again, this is far faster
than raster-scanning, which requires at least two $45$ second scans
for differencing.

\subsection{Video and Object Tracking}

\begin{figure}[h]
  \begin{centering}
    \includegraphics[scale=0.4]{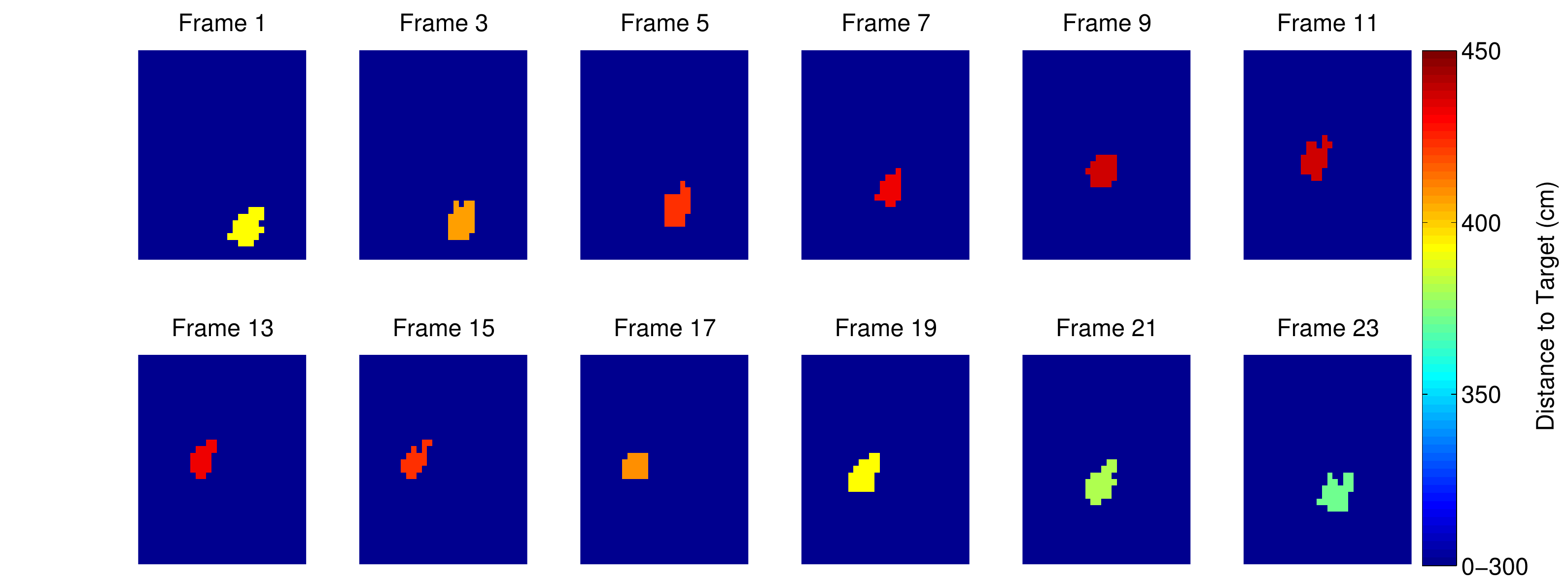}
    \caption{\textbf{Movie} Frames from a depth-map movie (\textcolor{blue}{Media 1}),
      of a three-dimensional pendulum consisting of a baseball
      suspended by a $170$ cm rope swinging through a $25$ degree
      solid angle.  The transverse resolution is $32\times 32$ pixels
      with a frame rate of $14$ frames per second.}
    \label{fig:frames}
  \end{centering}
\end{figure}

\begin{figure}[h]
  \begin{centering}
    \includegraphics[scale=0.45]{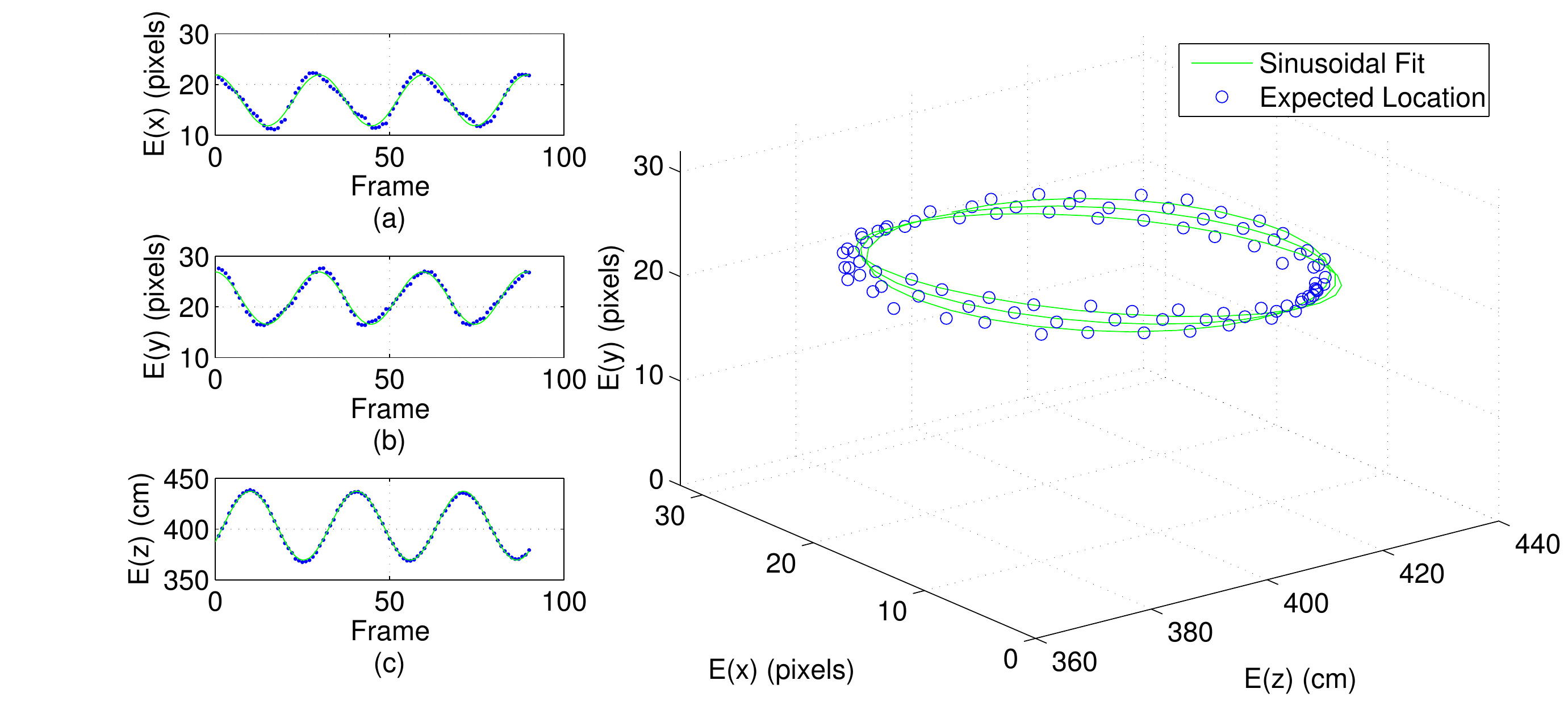}
    \caption{\textbf{Object Tracking} The expected values for
      transverse (x,y) and range (z) coordinates are given in (a),
      (b), and (c) as a function of frame number.  Blue circles show
      expected values obtained from reconstructed depth maps, while
      green lines give sinusoidal fits. (d) shows a 3D parametric plot
      of the pendulum's trajectory.}
    \label{fig:traj}
  \end{centering}
\end{figure}

At lower spatial resolution, the system is capable of video and
object tracking. To demonstrate this, we acquired video at $32\times
32$ pixel resolution of a three-dimensional pendulum consisting of a
baseball suspended from the ceiling by a rope. The lever arm was $170$
cm and the pendulum oscillated through a solid angle of approximately
$25$ degrees. The three dimensions are not oscillating in phase, so
the trajectory is elliptical.

Movie frames were acquired with $m=0.1n=99$ with a per-pattern dwell
time $t_p=1/1440$ sec. Each frame required $.07$ sec to acquire for a
frame rate slightly exceeding $14$ frames per second. The sparsity
promoter was $TV$.  Sample frames showing the pendulum moving through
a single period are given in Fig. \ref{fig:frames}, where alternate
frames are skipped for presentation. We clearly recover images of the
pendulum oscillating in all three dimensions. The full movie can be
viewed online (\textcolor{blue}{Media 1}).

The pendulum's position can be tracked by taking expected values for
its transverse location and averaging over its range to determine a
to-target distance. Fig. \ref{fig:traj} shows the computed pendulum
trajectory. Figs \ref{fig:traj}(a), \ref{fig:traj}(b) and
\ref{fig:traj}(c) give the expected $x$, $y$ and $z$ values for
pendulum location as a function of frame number. These can be combined
to yield the three-dimensional, parametric trajectory given in
Fig. \ref{fig:traj}(d). Sinusoidal fits are in good agreement with the
expected values, particularly in the depth dimension.

\section{Protocol Trade-offs and Limitations}

The strength of our protocol lies in its simplicity; the well-known
single-pixel camera can be adapted for ranging by simply adding a
pulsed source. All the standard, linear CS techniques are otherwise
used. This is because our protocol treats $X_Q$ as an image
itself. Our technique is also very natural for photon counting as
per-photon TOF is easy to measure and sum. However, this ease-of-use
does come with some potential trade-offs.

Our protocol does not consider sparsity in the depth map itself. Depth
maps have been shown to generally be sparser than intensity images
\cite{Kirmani:2011}. Protocols which directly recover the depth map,
such as in Ref. \cite{Kirmani:2011}, can leverage this sparsity to
require fewer measurements. Because incoherent measurements are
nonlinear in the depth map, this comes at the cost of more complex
measurement and reconstruction schemes.

In our protocol, $X_Q$ is treated like an image. This is reasonable
because it is simply a scaling of the intensity map by the depth map;
it maintains much of the structure of an image and still qualitatively
appears like an image to a viewer. Because $X_Q$ is treated like an
image, we use sparsity maximizers common for images such as wavelets
and total variation.  In these representations, $X_Q$ is likely to be
more complex (less sparse) than $X_I$, so the complexity of $X_Q$ is a
limiting factor reducing the number of measurements. It is possible
that other representations may be more appropriate for $X_Q$. A full
analysis of the complexity of $X_Q$ and its absolute best
representation remains an open problem.

Characteristics of the laser pulse, such as its width and shape, are
also not included in the protocol. Incorporating these in the
optimization may improve range resolution at the cost of a more
complex objective function. The current system can accurately
discriminate between objects placed less than a pulse width apart, but
this discrimination breaks down as the relative photon TOF between
objects becomes ambiguous. This is acceptable for many ranging tasks,
but very high range resolution depth mapping may require improvements.

Finally, our objective Eq. (\ref{eq:objective}) uses the standard least
squares penalty to ensure reconstructions are consistent with
measurements. At very low photon numbers, where the Poissonian photon
number distribution deviates markedly from Gaussian statistics, a
logarithmic penalty gives superior performance but requires different
solvers \cite{willett:2009}. Data sets used in this manuscript
typically averaged thousands of photons per measurement, so this was a
non-issue. This change would be critical if photon numbers approached
single-digits per measurement.

Ultimately, increased complexity in the protocol must be weighed
against potentially diminishing returns. In its current form, our
protocol demonstrates most of the benefits CS provides over standard
techniques. We do not claim the scheme is fundamentally optimal;
rather it is a practical approach that also provides a new perspective
on the compressive depth mapping problem.

\section{Conclusion}

We demonstrate a simple protocol for extending the single-pixel camera
to time-of-flight depth mapping with a photon counting detector. Our
approach requires only linear measurements so standard compressive
sensing techniques can be used. The photon-counting detector allows us
to work at sub-picowatt light levels. We show imaging $256\times 256$
pixel resolution with practical acquisition times. By differencing
random projections of two instances of a scene, we directly recover
novelty images to efficiently remove static clutter and show only
changes in a scene. We also acquire $32\times 32$ pixel video at $14$
frames per second to enable accurate three-dimensional object
tracking. Our system is much easier to scale to high
spatial-resolution than existing photon-counting lidar systems without
increased cost or complexity.

Single-pixel camera type devices are very flexible because operational
parameters can easily be changed just by switching detectors. While
our proof-of-principle system operates in the lab at $780$ nm, it
could adapt to infrared by switching to InGaAs APDs or superconducting
nano-wire detectors \cite{mccarthy:2013}. If faster acquisition times
are needed, one could use linear mode detectors or sum the output from
multiple photon-counting detector. Hyperspectral imaging is possible
with a frequency sensitive detector. Our protocol is a simple,
effective way to add high resolution depth mapping to many low-light
sensing applications.

%\section{Acknowledgments}

We acknowledge support from AFOSR grant FA9550-13-1-0019 and
DARPA DSO InPho grant W911NF-10-1-0404.

\end{document}